\documentclass[11pt,twoside]{article}
\usepackage{asp2006}
\usepackage{epsf}
\usepackage{graphics}
\usepackage{lscape}
\def\edcomment#1{\iffalse\marginpar{\raggedright\sl#1\/}\else\relax\fi}
\reversemarginpar

\begin{document}
\title{On the Polar Field Distribution as Observed by SOLIS}
\author{N.-E. Raouafi, J. W. Harvey and C. J. Henney}
\affil{National Solar Observatory, 950 N. Cherry Avenue, Tucson, AZ 85719, USA}

\begin{abstract} We use Vector Spectromagnetograph (VSM) chromospheric full-disk magnetograms, from
the Synoptic Optical Long-term Investigations of the Sun (SOLIS) project, to study the distribution
of magnetic field flux concentrations within the polar caps. We find that magnetic flux elements
preferentially appear toward lower latitudes within the polar caps away from the poles. This has
implications on numerous solar phenomena such as the formation and evolution of fine polar coronal
structures (i.e., polar plumes). Our results also have implications for the processes carrying the
magnetic flux from low to high latitudes (e.g., meridional circulation). \end{abstract}

\section{Introduction}

Near minima of the activity cycle of the Sun, solar poles are characterized by unipolar magnetic
dominated areas, in particular at photospheric heights (Babcock \& Livingston 1958; Babcock 1959;
Howard 1972; Timothy, Krieger, \& Vaiana 1975; Harvey, Harvey, \& Sheeley 1982; Varsik, Wilson, \&
Li 1999). The so-called ``polar caps'' are quite prominent features due to their large sizes and
their continuous presence through the solar cycle. They are the counterpart of the coronal polar
holes where the fast solar wind streams (Schwenn, Muelhaeuser, \& Rosenbauer 1979) and plasma
heating (Schatzman 1949; Osterbrock 1961) originate. The physical processes laying behind these
phenomena are still far from being completely understood. A significant portion of the mostly
unipolar magnetic flux opens increasingly with height and fills most the heliosphere that
contributes to the formation of the heliospheric magnetic field.

The solar polar caps are not sufficiently studied for different reasons (observational and
modeling). In addition to the important projection effects due to the solar geometry and the lack
of off-ecliptic-plane observational facilities of the magnetic field, the latter is intrinsically
weak close to the solar poles (Grigor'ev 1988; Bogart, Hoeksema, \& Scherrer 1992; Varsik, Wilson,
\& Li 1999; Varsik et al. 2002). In addition, the solar poles are only partially observable from
the ecliptic plane due to $B_0$. However, instrumental limitations in matter of sensitivity  are
the major problem for accurately measuring and studying the polar fields. 

Complex dynamics in the solar upper convection zone together with other dynamical phenomena (e.g.,
differential rotation), magnetic fields of decaying active regions drift monotonously to form the
large-scale field observed on the solar surface, in particular the polar caps (see Leighton 1964;
DeVore \& Sheeley 1987; Sheeley, Nash, \& Wang 1987; Bilenko 2002). The same processes contributes
significantly to the so-called polar magnetic reversal (Babcock \& Babcock 1955) that is the change
of the dominant magnetic polarity to the opposite from a solar cycle to the following one (Bilenko
2002). Under the effect of solar differential rotation, supergranular diffusion and meridional
circulation, the poleward migration of the active region trailing polarities (in relative excess
due to the diffusion of the leading polarity of emerging bipolar regions near solar minimum across
the solar equator as suggested by Wang, Nash, \& Sheeley 1989) has been suggested to be responsible
of the reversal of the magnetic polarity of the solar poles (e.g., Babcock \& Babcock 1955; Babcock
1961; Leighton 1964). Meridional flows has been observed and identified as the main process of flux
transport from low to upper latitudes (Howard 1974; Duvall 1979; Cameron \& Hopkins 1998; Snodgrass
\& Dailey 1996; Durrant, Turner, \& Wilson 2004) through helioseismology and magnetic tracers
(Duvall 1979; LaBonte \& Howard 1982; Ulrich et al. 1988; Howard \& LaBonte 1981; Topka et al.
1982; Wang, Nash, \& Sheeley 1989). The weak flow, of few times 10~m~s$^{-1}$ at best, can be
measured only to mid solar latitudes since the techniques used fails above roughly
$45^\circ-50^\circ$. However, it is of capital importance to obtain additional constraints on these
flows up to high latitudes close to the solar poles. This would be of great use for models dealing
with flux transport and the solar dynamo.

Raouafi, Harvey, \& Solanki (2006a,b; 2007) studied the plasma dynamic properties in the coronal
polar plumes by means of forward modeling. They found that modeled spectroscopic emissions from
polar coronal plumes rooted close to the solar poles do not match the observed coronal spectra by
the UltraViolet Coronagraph Spectrometer (UVCS; Kohl et al. 1995) on board the Solar and
Heliospheric Observatory (SOHO; Domingo, Fleck, \& Poland 1995). They speculated that polar plumes
would preferentially be based within the polar holes about 10$^\circ$~away from the solar poles. It
is well known from previous studies that polar plumes are the product of magnetic reconnection of a
relatively large unipolar flux element with an emerging opposite polarity. Thus, the study of the
polar magnetic flux distribution would yield insights and clues about the distribution of the
coronal structures, such as polar plumes. In the present paper, we confirm Raouafi, Harvey, \&
Solanki's results through the study of the polar flux distribution as a function of latitude and
suggest the probable effects on solar phenomena that are at the origin of such a distribution.

Saito (1958) used eclipse observations and found a similar distribution of the so-called ``polar
rays''.

\section{OBSERVATIONS AND DATA ANALYSIS}

 \begin{figure}[!h]
 \plotone{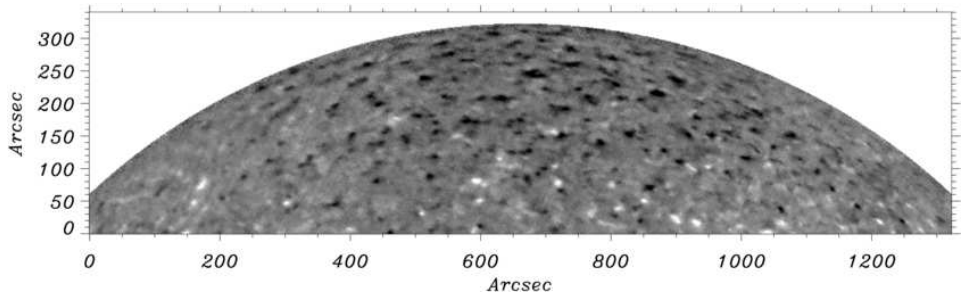}  
 \caption{Typical chromospheric (Ca~{\sc{ii}}~854.2~nm) LOS-magnetogram from the SOLIS/VSM recorded on
October 22, 2006. Note the good visibility  of magnetic flux elements everywhere on the solar
disk, in particular in the high latitude polar caps.
     \label{solis_8542_071206_fulldisk_sw24_paper}}
 \end{figure}

The magnetic line-of-sight (LOS) component obtained from the polarization of photospheric lines is
not sufficiently strong and does not allow to study accurately the distribution of magnetic flux
near the poles. In addition, the transverse field measurements are not sufficiently sensitive.
These make the task of addressing polar fields from photospheric magnetograms rather difficult.
However, LOS-magnetograms obtained from chromospheric lines (e.g., Ca~{\sc{ii}}~854.2~nm) benefit
from the canopy structure of the magnetic field which yield strong signals to the solar limb (see
Figure~\ref{solis_8542_071206_fulldisk_sw24_paper}). It is noteworthy that the photospheric
LOS-magnetograms from the same instrument are of equally good quality but near the limb where the
flux signal is mixed with a dynamic horizontal field (Harvey et al. 2007) that is not as strong in
the chromosphere.

Figure~\ref{solis_8542_071206_fulldisk_sw24_paper} illustrates a typical VSM LOS-magnetogram in the
Ca~{\sc{ii}}~8542~{\AA} line recorded on October 22$^{\rm{nd}}$, 2006. Note the visibility of the
magnetic elements demonstrating the capability of the SOLIS-VSM instrument to sense the magnetic
field everywhere on the solar disk. It is noteworthy that the high signal to noise in the polar
regions up to the polar limb makes SOLIS a suitable instrument to study polar fields in the absence
of out-of-ecliptic measurements.

The chromospheric (Ca~{\sc{ii}}~8542~{\AA}) LOS-magnetograms from SOLIS-VSM are utilized to study
the latitude distribution of the unsigned magnetic flux in the polar regions. We are interested in
the distribution of magnetic flux elements regardless of their polarities. In addition, polar caps
are typically dominated by one polarity each which also justifies the use of unsigned magnetograms.
The opposite polarity in these areas contributes little to the total flux and is represented by
relatively small-diffuse flux elements.

The non-uniform visibility across the solar disk is corrected for assuming an East-West uniform
distribution of magnetic flux elements when a considerable number of magnetograms are averaged. This
is more reliable for quiet sun regions. The correction is dependent on the distance to
disk center and is represented by
\begin{equation}
f=C_0\;\left(\frac{r}{R_{\sun}}\right)^n+1,
\end{equation}
where $C_0$ is an adjustable parameter and is found to be nearly equal to 1.1 and $r$ is the
distance to disk center. This empirical correction enhances magnetic fields close the solar limb
that are relatively noisy. By setting all magnetic fields with strength below a given threshold,
that is equal to 5 Gauss in the present case, to zero, the noise effect is reduced significantly.
In addition, thresholding the magnetograms allow for the boundaries of magnetic flux elements to be
easily defined.

 \begin{figure}[!h]
 \plotone{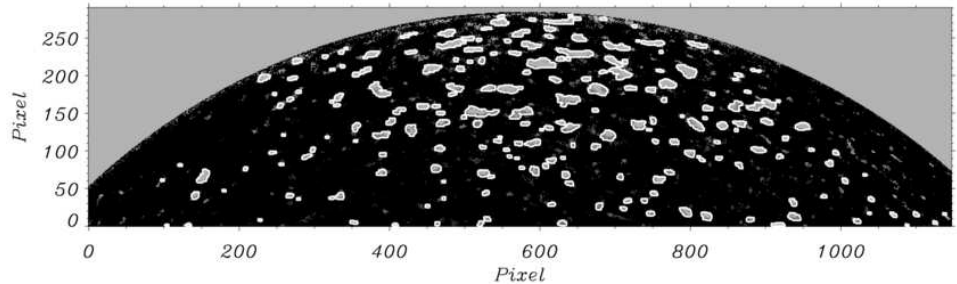}  
 \caption{Flux
elements selected by the method we use here to study the latitude flux distribution. The minimum
size of flux element to be chosen in the present case is $5\times3$ pixel.
Magnetic fields below 5 Gauss were set to zero before the radial correction is performed.
\label{Flux_count_paper_x5y3_sw24}}
 \end{figure}

A size and shape dependent recognition methods of closed structures is used to identify and locate
the flux elements. The method dependence on the size and shape of flux elements allows for studying
the distribution of different population of magnetic structures.
Figure~\ref{Flux_count_paper_x5y3_sw24} illustrates how the used method for the selection of
magnetic elements works with a size threshold of $5\times3$ pixels (1 pixel $\approx$ 1.13 arcsec).
Note that magnetic elements with smaller sizes are not selected. The distribution of the selected
elements as a function of latitude is obtained by determining the an average location for every
selected feature. However, the obtained distribution is absolute and might be biased by the
latitude area dependence. In order to stay clear of that, the obtained distribution are normalized
by the latitude area distribution taking into account the solar geometry (i.e., $B_0$ angle). Since
single histograms do not show clearly the distribution of magnetic flux elements due to statistical
reasons, they are monthly averaged. 

\section{RESULTS AND DISCUSSION}

 \begin{figure}[!h]
 \plotone{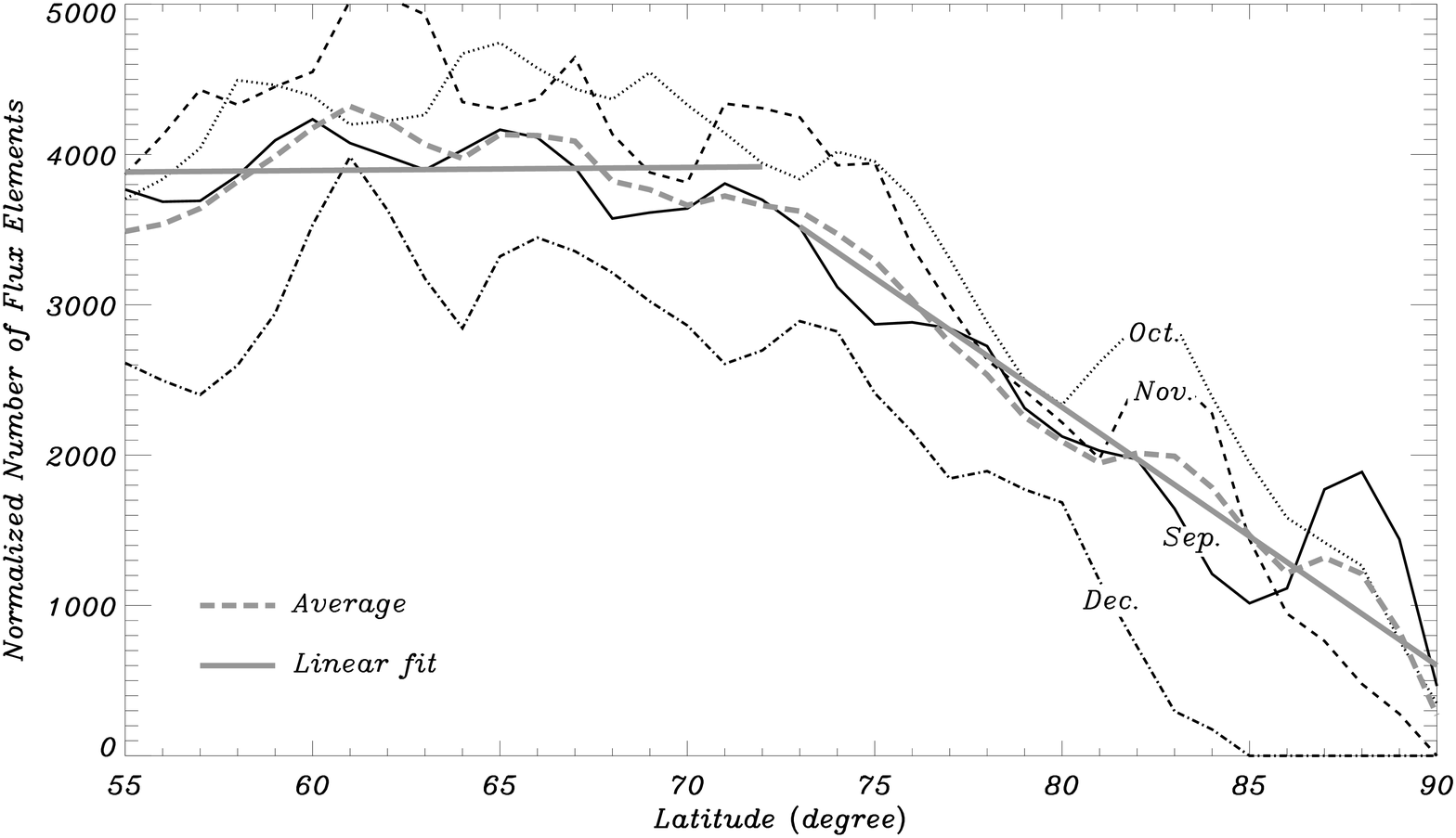} 
 \caption{Monthly averaged histograms of the magnetic flux distribution for features larger than $5\times3$
pixel as a function of latitude for the north polar cap. The grey
dashed curve is the average of the four previous ones. The straight lines are linear fits for
portions of the grey dashed curves for latitudes ranging from $55^\circ-73^\circ$ and $74^\circ-90^\circ$.
\label{svsm_m2100_S2_200609_12_X5y3_sw24}}
 \end{figure}

Figure~\ref{svsm_m2100_S2_200609_12_X5y3_sw24} displays the monthly averaged distributions of
magnetic flux for the north polar cap as a function of latitude from September till December 2006.
The different curves are indexed accordingly in the same figure and the average of the obtained
monthly histograms is given by the grey dashed line. The approximately horizontal and oblique
straight lines are linear fits of portions of the latter for latitudes ranging from
$55^\circ-73^\circ$ and $74^\circ-90^\circ$, respectively. These distributions are obtained by
imposing a minimum size of $5\times3$ pixel as shown by
Figure~\ref{Flux_count_paper_x5y3_sw24}.

The linear fits show that the distribution density of polar flux elements normalized by the surface
area is relatively uniform at low latitudes up to approximately $70^\circ-75^\circ$. At higher
latitudes this density has a decreasing trend toward the solar pole. It is clear from curves that
the decrease is persistent and from the linear fit that it is significant although few peaks shows
up very close to the pole (see the September curve for instance). This suggests that flux elements,
in particular relatively large ones (see Raouafi, Harvey, \& Henney 2007), are more concentrated at
low latitudes rather than higher latitude elements within the polar cap.

The flux distribution as a function of latitude has an impact on numerous solar phenomena such the
location of fine coronal polar structures. Raouafi, Harvey, \& Solanki (2007) built different
numerical models to study the plasma dynamics in polar plumes. In all these models, that cover
almost all the possible cases for the outflow velocity and turbulence of the plasma, they found
that polar plumes rooted close to the pole are found to have spectral signatures that are not
observed in the solar corona. They concluded that polar plumes would preferentially be rooted more
that about 10$^\circ$ from the pole. There is also observational evidence that the base of polar
plumes, in particular bright ones, are relatively large magnetic flux areas dominated by one
polarity (Saito \& Tanaka (1957a,b); Harvey 1965; Newkirk \& Harvey 1968; Lindblom 1990; Allen
1994). The latter and Raouafi, Harvey, \& Solanki's findings would agree well with the picture we
are drawing here on the density distribution of magnetic flux on the polar caps. This is more
evident in the case where we consider relatively large flux elements.

The results of the present study are of capital importance for solar phenomena that are
inaccurately determined at high latitudes, such as meridional circulation. Meridional flows are
widely believed to be one of the main factors causing the inversion of the magnetic polarity of the
poles from one solar cycle to the next. Together with supergranular diffusion, meridional flows
bring the magnetic flux of decaying active regions from low to high latitudes causing then the well
known polar magnetic reversal. Due to the flow signal weakness in Doppler measurements (Zhao \&
Kosovichev 2004; Haber et al. 2002) it is not well understood how this phenomenon operates at high
latitudes and in particular close to the poles. The density distribution of the unsigned magnetic
flux at the polar regions reported on here suggests that the mechanisms responsible for the flux
transport greatly diminish before reaching the poles. Such results would be of great importance for
models dealing with flux transport and the solar dynamo.

\section{CONCLUSION}

We studied the distribution of magnetic flux elements as a function of latitude near solar minimum
of activity utilizing LOS-magnetograms from the SOLIS-VSM magnetograph. Chromospheric magnetograms
are preferred to photospheric ones because of the better signal in the LOS magnetic component due
the canopy structure of the magnetic field. Additional effects, such as the seething in the
horizontal photospheric magnetic fields, are avoided by using chromospheric magnetograms. The
magnetic data is recorded almost daily from September through December 2006. The north pole data
was chosen for two reasons:
\begin{itemize}
\item[$\bullet$] the north polar cap was better developed than the southern one during this period
of time;
\item[$\bullet$] the north pole was more visible than the southern one ($B_0>0$).
\end{itemize}

A recognition method of closed structures has been used to select magnetic flux elements.
Thresholds on size and shape of the elements to be selected are tunable parameters of the method.
The average spherical coordinates (latitude and longitude) of the selected elements are computed
and their daily latitudinal distributions are determined. In order to remove any effect of the
surface area of the polar cap, the histograms are normalized with respect to the area corresponding
to each latitude bin taking into account the different solar geometrical parameters for the each
magnetogram. Since daily distributions do not reflect the real density of magnetic elements for
statistical reasons, monthly averages of the normalized histograms are then computed.

The density of the unsigned magnetic flux elements within the polar magnetic cap (normalized by the
surface area of the polar cap) is found to be strongly dependent on solar latitude. Flux elements
are found relatively uniformly dense up to latitudes of about $75^\circ-80^\circ$ followed by a
significant decrease in their latitudinal distribution at higher latitudes close to the solar pole.
Although, a difference in the distribution of relatively small and large flux elements is found,
the density decreasing behavior toward the pole is found in all the data considered (for details
see Raouafi, Harvey, \& Henney 2007). We believe that the mechanisms responsible for the transport
of the magnetic flux of low latitude decaying active regions toward high polar latitudes are less
efficient close to the poles. This means that the meridional circulation responsible for the flux
transport slows before reaching the poles. Such a result would have a significant impact on the
theories and models dealing with flux transport. These results also put important constraints on
solar phenomena that are inaccurately (if at all) determined by other means, such as helioseismic
studies that become inefficient at high latitudes.

\acknowledgments
SOLIS data used here are produced cooperatively  by NSF/NSO and NASA/LWS. The National Solar
Observatory (NSO) is operated by the Association of Universities for Research in Astronomy, Inc.,
under cooperative agreement with the National Science Foundation. NER's work is supported by NSO
and NASA grant NNH05AA12I.

\end{document}